%% file: main.tex
\documentclass{article}

\usepackage[final]{nips_2018}
\usepackage[utf8]{inputenc}
\usepackage[T1]{fontenc}
\usepackage{hyperref}
\usepackage{url}
\usepackage{booktabs}
\usepackage{amsmath}
\usepackage{amsfonts}
\usepackage{nicefrac}
\usepackage{microtype}
\usepackage{hyperref}
\usepackage[pdftex, dvipsnames]{xcolor}
\usepackage[pdftex]{graphicx}
\usepackage{enumitem}

\title{Idiosyncrasies and challenges of data driven learning in electronic trading}

\author{
	\begin{tabular}{cc}
		Vangelis Bacoyannis & Vacslav Glukhov \\
		\texttt{\href{mailto:vangelis.bacoyannis@jpmorgan.com}{vangelis.bacoyannis@jpmorgan.com}} & \texttt{\href{mailto:vacslav.glukhov@jpmorgan.com}{vacslav.glukhov@jpmorgan.com}} \\
		\\
		Tom Jin & Jonathan Kochems \\
		\texttt{\href{mailto:tom.jin@jpmorgan.com}{tom.jin@jpmorgan.com}} & 
		\texttt{\href{mailto:jonathan.a.kochems@jpmorgan.com}{jonathan.a.kochems@jpmorgan.com}} \\
		\\
		\multicolumn{2}{c}{Doo Re Song} \\
		\multicolumn{2}{c}{\texttt{\href{mailto:doore.song@jpmorgan.com}{doore.song@jpmorgan.com}}}
	\end{tabular}
	}

\begin{document}

\maketitle

\begin{abstract}
	We outline the idiosyncrasies of neural information processing and machine learning in quantitative finance.  We also present some of the approaches we take towards solving the fundamental challenges we face. 
	\footnote{\tiny This Material has been prepared by J.P. Morgan's EMEA eTrading Quantitative Solutions Team. Opinions and estimates constitute our judgement as at the date of this \textbf{Material}, are provided for your informational purposes only and are subject to change without notice. Neither J.P. Morgan Securities plc nor its affiliates and / or subsidiaries (collectively \textbf{J.P. Morgan}) warrant its completeness or accuracy. This Material is not the product of J.P. Morgan's Research Department and therefore, has not been prepared in accordance with legal requirements to promote the independence of research, including but not limited to, the prohibition on the dealing ahead of the dissemination of investment research. It is not a research report and is not intended as such. This Material is not intended as research, a recommendation, advice, offer or solicitation for the purchase or sale of any financial product or service, or to be used in any way for evaluating the merits of participating in any transaction. Please consult your own advisors regarding legal, tax, accounting or any other aspects including suitability implications for your particular circumstances.  J.P. Morgan disclaims any responsibility or liability whatsoever for the quality, accuracy or completeness of the information herein, and for any reliance on, or use of this material in any way. Important disclosures at: \href{https://www.jpmorgan.com/disclosures}{www.jpmorgan.com/disclosures}. \textcopyright 2018 JPMorgan Chase \& Co. All rights reserved.}
\end{abstract}

\input{section_intro}
\input{section_cultures}
\input{section_dimension}
\input{section_policy}
\input{section_compute}
\input{section_uncertainty}

\bibliographystyle{plainnat}
\small
\bibliography{references}

\end{document}

%% file: section_intro.tex
\section{Introduction}

Portfolios of financial instruments held by pension funds and other asset managers undergo periodic rebalances, sometimes radical. Agency electronic trading, a service provided by brokers such as big banks and specialized broker companies, helps make these transitions efficient. Savings provided by efficient portfolio transitions are passed back to the clients, and, in turn, to the ultimate beneficiaries of these portfolios  --- teachers, doctors, firefighters, government employees, workers, hedge fund operators, etc.  

\begin{figure}[h]
\centering
\includegraphics[width=0.8\linewidth]{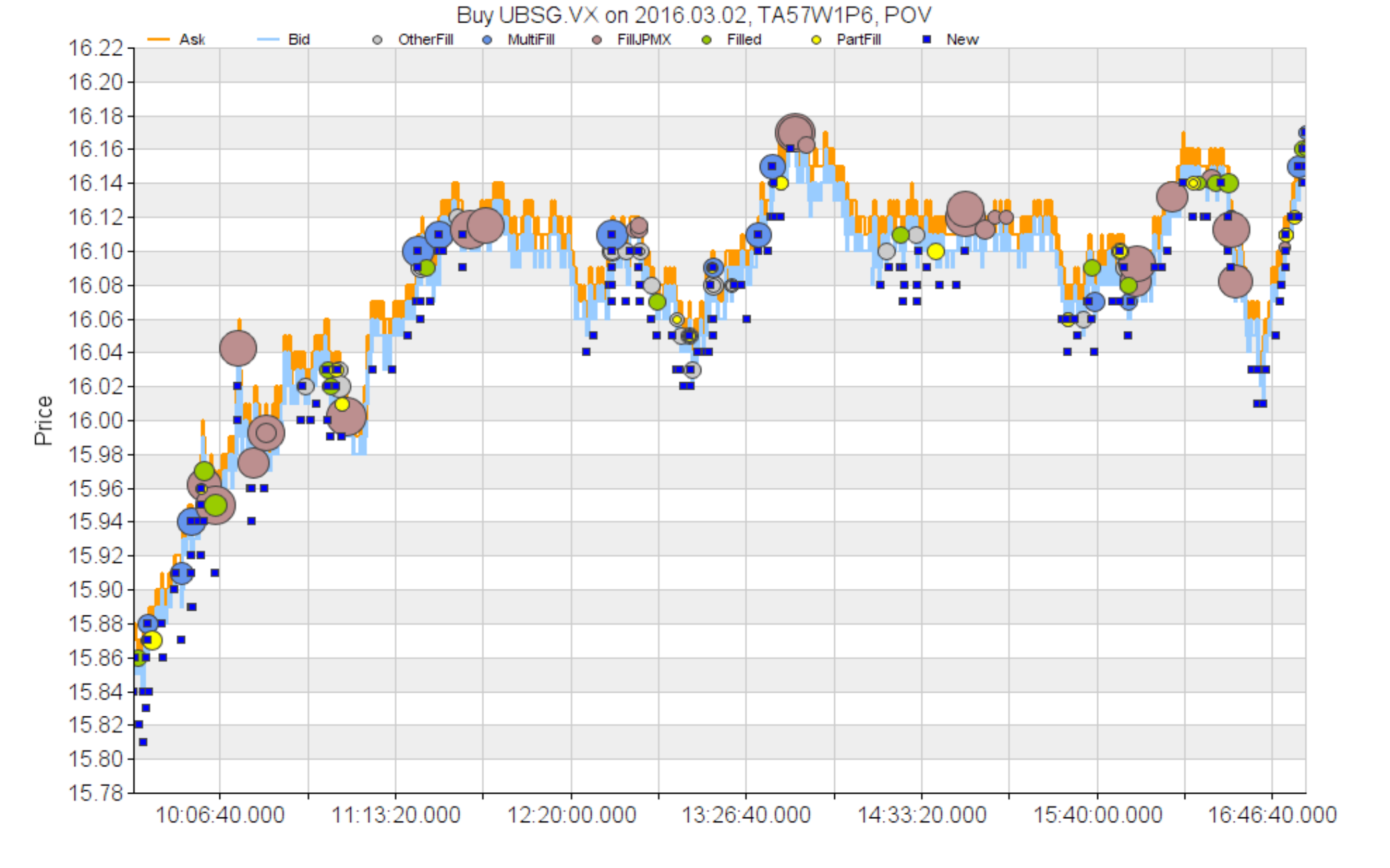}
\caption{Percentage of Volume (PoV) algorithm in action: deep blue - passive orders, light blue and orange: bid and ask market prices, circles - order fills}
\label{fig:pov_example}
\end{figure}

The globalization of asset trading, the emergence of ultrafast information technology and lightning fast communications made it impossible for humans to efficiently compete in the routine low-level decision making process. Today most micro-level trading decisions in equities and electronic future contracts are made by \emph{algorithms}: they define where to trade, at what price, and what quantity.  An example of the algorithm in action is in Fig.~\ref{fig:pov_example}.

Given their overarching investment and execution objectives, clients typically transmit specific instructions with constraints and preferences to the execution broker. To give just a few examples, clients may want to preserve currency neutrality in their portfolio transitions, so that the amount sold is roughly equal to the amount bought. Clients can also express their risk preferences and specify that the executed basket of securities is exposed in a controlled way to certain sectors, countries or industries. For single order executions clients may want to control how the execution of the order affects market price (control market impact), or control how the order is exposed to market volatility (control risk), or specify an urgency level to optimally balance  both market impact and risk.

In order to fulfil these multifaceted and sometimes conflicting objectives electronic trading algorithms operate on multiple levels of granularity. Making decisions on every level is informed by market analytics and quantitative models. Traditionally, electronic trading algorithms were a blend of scientific, quantitative models which expressed quantitative views of how the world works, and rules and heuristics which expressed practical experience, observations and preferences of human traders and users of algorithms.  The logic of a traditional trading algorithm and its accompanying models  is often encapsulated in tens of thousands lines of hand-written, hard to maintain and modify code. Responding to clients' objectives and changes in financial markets, human-coded algorithms tend to suffer from ``feature creep'' and eventually accumulate many layers of logic, parameters,  and tweaks to handle special cases. 
 
The financial services industry is heavily regulated. In some regions very specific requirements, such as the concept of ``best execution'' in EMEA \citep{MiFIDII}, are placed on the participants. Conforming to these requirements and achieving efficiency of algorithmic trading is challenging: changing market conditions and market structure, regulatory constraints, and clients' multiple objectives and preferences make the design and development of electronic trading algorithms a daunting task. 
The possibility of using data-centric approaches, neural processing, and machine learning presents an attractive opportunity to streamline the development and improve the efficiency of applications in electronic trading business.

In this short paper we attempt to bridge the existing methodological gap between academia and the financial industry. We present practical challenges and idiosyncrasies which arise in electronic trading which we hope will be inspirational for academic researchers. 

%% file: section_cultures.tex
\section{Three cultures of data-centric applications in quantitative finance}

In this section we first follow and then take further the argument developed by Peter Norvig in~\cite{PN:2011}.
The following three cultures are  associated with the three consecutive generational waves of researchers in the field. 

\subsection{Data modelling culture} 
This culture is characterized by a belief that nature (and financial markets) can be described as a black box with a relatively simple model inside which actually generates the observational data.
The task of quantitative finance is to find a plausible functional approximation for this data generating process, a quantitative model, and to extract its parameters  from the data.
The output of the model is then fed into  quantitative  decision-making processes. Complexity of markets and behaviours of market participant present the main challenge to the data modelling culture: simple models do not necessarily capture all essential properties of the environment.
One can argue that simple models often give a false sense of certainty, and for this reason are prone to abject failures. 

\subsection{Machine learning culture} 
For the machine learning culture an agnostic approach is taken to the question whether nature and financial markets are simple.
We do have good reason to suspect that it is not: empirically the world of finance looks more Darwinian than Newtonian: it is constantly evolving, and observed processes including trading in electronic markets are best described as emerging behaviours rather than data generating machines.
In the machine learning culture complex and sometimes opaque functions are used to model the observations.  Researchers don't claim that these functions reveal the nature of the underlying processes.
As in the data modelling culture,  machine learning models are built and their output is fed into decision-making processes.
Complex models are prone to failures as well: risk of the model failure increases with its complexity. 

\subsection{Algorithmic decision-making culture} 

Here our focus is on decision-making rather than on model-building.
We bypass the stage of learning “how the world works” and proceed directly to training electronic \emph{agents} to distinguish good decisions from bad decisions.
The challenge presented by this approach is in our ability to understand and explain the decisions the algorithmic agent takes, to make sense of its policies, and to be able to ensure that the agent produces sensible actions in all, including hypothetical, environments.
In the algorithmic decision-making culture the agent learns that certain actions are bad because they lead to negative outcomes (\textit{malum in se}).
But we still have to inject values and rules and constraints that steer the agent away from taking actions which we view as prohibited (\textit{malum prohibitum}) but which the agent cannot learn from its environment and history. 

In this paper we show the interplay between the agent's constraints and rewards in one practical application of reinforcement learning.
We will also give an overview of specific challenges and how we tackle them using  computational resources and the many achievements of other AI teams across many industries and in academia.

%% file: section_dimension.tex
\section{Low to High Dimensionality and Back Again}

\subsection{High level decision-making}

From a very high level perspective, it is obvious that for every order there is an optimal execution rate or execution schedule, that is, speed with which order is executed, or the duration of its execution in the marketplace.

First, an order of almost any size can be executed instantaneously --- if the client is insensitive to the cost of execution and is willing to pay the price. No doubt such execution is unreasonable, inefficient  and potentially prohibitively expensive under normal circumstances. Such execution would, with high probability, affect market prices. 

On the other hand, a parent order can exert almost no pressure on markets if it is executed with child orders at an infinitely slow rate. Such execution is unreasonable, too, for no client is insensitive to the \emph{possibility} of undisturbed market prices going against the order (up for a buy order, down for the sell order). The longer the execution, the higher the probability of market prices going against the best interest of the client, that is, the higher the risk. 

From this simple consideration of the two limiting cases it is easy to see that there must be an optimal rate of execution or an optimal execution schedule. It is also easy to see how the client's preferences and tolerances come into play: the efficient rate is determined by the client's tolerance to market impact and appetite for risk. This is an example of high-level decision-making  under uncertainty  informed by high-level analytics and quantitative models. 

This also illustrates the important truth we often discover and rediscover in electronic trading and elsewhere in quantitative finance: there are no solutions, only trade-offs.  

\subsection{Low level decision-making}

Once a rough optimal rate or schedule is found, the next level of decision-making deals with the implementation of the schedule.  In order to stay on schedule, the agent typically tries to blend with the rest of the market: being an outlier is penalized because it reveals the agent's intention.  The agent creates marketplace orders which mimic other  participants' orders  --- both in size and in prices. 

It is here where we find the dimensionality explosion. 

Describing the market state of the limit order book is a variable dimension and high dimension problem. Each price level is a queue of differently sized orders from different market participants. These queues could be arbitrarily long or empty. At any particular time the most important levels are those which correspond to the prevailing bid and ask prices. However, there is a significant volume of orders at deeper levels and speculative far away levels.
As trades occur and orders are received and withdrawn, the order book is in constant change. Every observed market state can potentially evolve into an almost infinite number of other market states. 

In this environment the set of feasible decisions, even considered on the most elementary level of the order time, price, size, and duration, is very large and dense. The agent has to decide at which price and what quantity to place and if desired make multiple orders at different prices or make additional orders at prices where we already have an order in place. If the price of an order is not at the market price then the order will remain the book indefinitely until the price reaches that point, if it does. This action space is necessarily dynamic and complex as placing orders at depth is necessary to achieve price improvement and gradually orders are filled with price-time priority from the order book. A final complication depending on the available venues for execution is that there may be multiple suitable trading venues  and order types. 

A game of Chess is about 40 steps long. A game of Go is about 200 steps long. If a medium frequency electronic trading algorithm reconsiders its options every second, it amounts to 3600 steps per hour. For Chess or Go, it is moving one piece among the eligible pieces and moves per pieces.

For electronic trading, an action is a collection of child orders: it consists of multiple concurrent orders with different characteristics: price, size, order type etc. For example, one action can simultaneously be submitting a passive buy order and an aggressive buy order. The passive child order will rest in the order book at the price specified and thus provide liquidity to other market participants. Providing liquidity might eventually be rewarded at the time of trade by locally capturing the spread: trading at a better price vs someone who makes the same trade by taking liquidity. The aggressive child order, on the other hand, can be sent out to capture an opportunity as anticipating a price move. Both form one action.  The resulting action space is massively large and increases exponentially with the number of combinations of characteristics we want to use at a moment in time.

It's not entirely clear how to define efficiency of each action. One can argue that efficiency and optimality of decision-making for an electronic trading agent can be in detecting and capturing opportunities (``good'' trades), and in avoiding pitfalls (``bad'' trades). The problem with this fine-grained definition is not only that many opportunities are short lived and exist possibly on a microsecond scale only. More important is the fact that whether the trade is going to be good or bad is not known with certainty until well after the trade is executed (or avoided). 

The consequence is that local optimality does not necessarily translate into a global optimality: what could be considered as a bad trade now could turn out to be an excellent trade by the end of the day. In that sense, we are as interested in exploring and redefining what an opportunity is as we are define how to act. We refer to this distinctive aspect of electronic trading as non-local optimality.

A possible  (but not necessarily unique or best) global objective for the agent is its  ability to blend with the rest of the market.  If this is the case, a reward function to achieve the best execution price relative to the volume weighted average price, can be used. The strategy has to find a balance between market impact from trading too quickly and moving the price, on one hand, and market risk, from exogenous price movements as a result of trading too slowly, on the other hand. A significant part of this problem is encapsulating the state information and action space in a manner suitable for to fit models and use machine learning methods. This involves summarising the market state with potentially huge, variable and frequently changing dimension and order state, both parent order and child orders outstanding for model inputs. Then selecting one of a variable number of actions in response.

\subsection{Prior work}

There is an interesting breadth of existing work in this area that in general approaches individual aspects of this problem. Some works include prior setups for reinforcement learning in a small dimension environment whilst others consider representing the data in a succinct and fixed dimensional manner. \cite{DBLP:journals/tsp/AkbarzadehTS18}~looks at this problem with a view to performing online learning to drive the algorithm. The performance is however constrained by only making market orders. 
 
\cite{DBLP:conf/icml/NevmyvakaFK06}~defined an entire reinforcement learning problem but this was severely restricted by an action space that admitted a single order where new orders cancel older ones. In~\cite{zhang2018deeplob} a limit order book was summarised into a 40 dimensional vector containing price and volume information from the 10 price levels either side of the spread. This information is normalised based on the previous day's trading and used to predict market movement. \cite{doering2017convolutional} goes further by designing 4 matrices that contain the order book, trades, new orders and order cancellations at the expense of quadrupling the dimensions and using particularly sparse data.

The future research directions primarily target the continued research and development into trading agents based on reinforcement learning methods. Core to this is effective dimensionality reduction to encapsulate as much information about the current market and the state of existing orders both of which need a fixed dimensional representation of highly variable dimension data. Existing methods simplify the order management process by assuming a fixed number of outstanding child orders at unique prices which is unduly restrictive compared to the actions available to a human trader.

\subsection{A nano-description of our approach}

We are now running the second generation of our RL-based limit order placement engine.  We successfully train a policy with a bounded action space. To tackle the issues we have just described we use hierarchical learning and multi-agent training which leverage the domain knowledge. We train local policies (e.g.\ how to place aggressive orders vs how to place a passive order) on local short term objectives which differ in their rewards, step and time horizon characteristics. These local policies are then combined, and longer term policies then learn how to combine the local policies. 

We also believe that inverse reinforcement learning is very promising: leveraging the massive history of rollouts of human and algo policies on financial markets in order to build local rewards is an active field of research.

%% file: section_policy.tex
\section{Beyond policy learning in development of AI for electronic trading}

\subsection{Policy learning algorithms}

The core objective of RL is to maximize the aggregated rewards which approximates the true business objective. Policy learning algorithms that optimises a parametrized action policy on this objective have been a main focus of RL research. Recent studies apply renowned policy learning algorithms to the electronic trading business~\citep{DBLP:journals/tsp/AkbarzadehTS18}~\citep{DBLP:conf/icml/NevmyvakaFK06}. We would like to introduce other aspects of RL that sit beyond what policy learning algorithms are capable of.

\subsection{Hierarchical decision making}
Real application of AI in the electronic trading is typically characterized by a long time horizon. Client orders take many minutes or even hours (sometimes days) while agents need to make decisions every few seconds or faster.  The time horizon issue extremely limits the agent's sampling frequency to far lower than what is necessary to fully integrate all available information about market dynamics.  

Furthermore, decision-making of the agent is time-inhomogeneous. Rather than being driven by the clock, it responds to the effects of its own actions as well as to substantial changes the environment. 

Temporal abstraction in RL therefore becomes a critical issue to cope with both a long temporal horizon and inhomogeneity of time. It is possible that the frame skipping metaphor– only making decision once every few time steps – is not applicable here. Semi-MDP (sMDP) has been a prominent venue to discover the temporal abstractive behaviour of RL agents~\citep{sutton1999between}. However, training a single policy for when to act and what to decide is still sample-inefficient. A possible solution is to couple sMDP with hierarchical RL (HRL). HRL is an approach where the decision model consists of layers of policies with different decision frequency from meta-policy to primitive policies.

Our formulation of an electronic trading agent is heavily motivated from Kulkarni's interpretation of rule-based deep HRL~\citep{DBLP:conf/nips/KulkarniNST16} since we can afford to impose reasonable rules  in order to construct meta-policy based on domain experiences. 
We also note the progress in the end-to-end (rule-free) hierarchical RL where the temporal abstractive property of meta-policy emerges from behaviour or goal clustering by primitive policies \citep{DBLP:conf/aaai/BaconHP17}\citep{DBLP:journals/corr/FoxKSG17}\citep{DBLP:conf/icml/VezhnevetsOSHJS17}. 

The core problems on the ability of AI agents to use temporal abstraction, however, remain unsolved: the agent's interpretation of sub-goals and intrinsic rewards in the context of overarching objectives, collapse of temporal abstraction at convergence, sample efficiency on exploration-heavy environments, and deep hierarchy.

\subsection{Algorithmic, regulatory and computational challenges} 
Electronic trading agents operate in a complex, evolving, and quickly changing environment.  Increased complexity of the agent which yields better decision-making  and improved efficiency can be a plus, but it might impact the agent's computational performance and  ultimately render infeasible deployment. 

Another constraint limiting the complexity of the agent  in agency electronic trading is the need to understand, to foresee, to explain its decisions --- from the highest level of decision making to the lowest. 

In certain regions it's a requirement that trading algorithms produce predictable, controllable, and explainable behaviours: the agents must not disrupt so called orderly market conditions, and the operator of the agent must be able to explain how the agent's actions produce best possible result for the client. 

A hierarchical approach helps here: it is based on the observation that the agents' decision can be separated into groups requiring different sampling frequencies and different levels of granularity. We  have already mentioned in the above that the hierarchical architecture and the HRL brings the possibility of separation of responsibility between  the agent's modules, and while we can still use neural processing and reinforcement learning in each of them, we are also able to manage the overall complexity of the agent, we can better understand what it does and why it does what it does.

%% file: section_compute.tex
\section{Hierarchical reinforcement learning scheme}

\subsection{Search-based optimization of meta-policy on simulation-heavy learning task}
Training an RL agent requires a number of episodic rollouts each of which cannot be parallelized due to the feedback loop between the agent and its environment. Gradient-based training of the agent suffers from a memory-heavy reservoir of experience pairs which are often redundant and noisy. Good behaviours are forgotten during the course of training unless the learning algorithm is strongly off-policy, while the success of gradient optimisation involving a moving objective is hardly guaranteed. For this reason pursuit of gradient-free optimization using parameter search algorithms is hence still a practical choice in spite of recent progress in policy learning algorithms.

We have earned substantial time efficiency by applying hyper-parameter optimization techniques to train parametrized agent with respect to episodic utility in full-scale of control~\citep{osborne2009gaussian}\citep{DBLP:conf/nips/BergstraBBK11}, which also improved overall execution performance without dealing with reward design. We would like to highlight the learning efficiency of parameter search algorithms. 

Computational constraints put limitations on using fully sequential optimization. We alleviate this by  exploring using less certain optimization with fewer sampled episodes per trial, but running it in parallel. Early-stopping of uninteresting  paths is a good compromise between the two. We hope, however, to continue this line of development with a Bayesian approach to early-stopping. 

\subsection{Scalable deep reinforcement learning for low-level decision processes}
In the previous section we mentioned some of the challenges we face with the development of electronic trading agents: 
the environment which is partially observable, the possible incommensurability of  time horizons between the fine-grained market dynamics, the agent's observations, and its overall business objective, the vast state space, and delayed and possibly staggered rewards.

As does every market participant, our agents change the environment in which they operate. We train agents in constructed simulated environments which attempt to reproduce some of the properties of real markets, but cannot currently reproduce all  of them. Particularly, we  strive to build a simulated environment which mimics real market's response to the agent's action.

\emph{Prima facie}, this demands an architecture which supports scalable simulations and scalable RL algorithms. 
The Gorila architecture~\citep{DBLP:journals/corr/NairSBAFMPSBPLM15} illustrates how the DQN algorithm~\citep{DBLP:journals/corr/MnihKSGAWR13} can be employed at scale yielding superior results. 
For A3C~\citep{DBLP:conf/icml/MnihBMGLHSK16}, a similar feat has been achieved recently by the IMPALA algorithm~\citep{DBLP:conf/icml/EspeholtSMSMWDF18}.
In general, it is an interesting question whether and how other RL algorithm schemes can be scaled to take advantage of large scale cluster compute in such a way as to obtain better performing policies. 
Evidence-based guidance would be very useful for practitioners who would like to exploit available compute resources for using a particular algorithm against their use-case. 

An exciting development is the emergence of open source RL frameworks such as OpenAI baselines~\citep{baselines}, ELF~\citep{DBLP:conf/nips/TianGSWZ17}, Horizon~\citep{gauci2018horizon}, dopamine~\citep{dopamine}, TRFL~\citep{trfl} and Ray RLlib~\citep{DBLP:journals/corr/abs-1712-05889}.
These frameworks and tools already make state-of-the-art reinforcement learning algorithms accessible to a much larger audience. 
However, the aforementioned RL frameworks are still young and nowhere near as mature and  ``production-ready'' as popular Deep Learning libraries such as Google TensorFlow, PyTorch, or Caffe. 
Having strong ecosystems and communities resembling the Deep Learning landscape around RL frameworks would be greatly conductive to expanding the accessibility of RL methods.

We found Ray RLlib useful. It is built from the ground up with distributed reinforcement learning in mind. 
Its foundation rests on a solid infrastructure which leverages task parallel and actor model~\citep{DBLP:books/mit/shriverW87/AghaH87} programming patterns, i.e.~programming paradigms which have proven to be very successful in designing efficient, large scale distributed computing systems~\citep{DBLP:journals/cacm/Armstrong10}.

RL experiments can be very time consuming and often complete in a sequence of partial experiments, sometimes interrupted by faults. Ray's design~\citep{DBLP:journals/corr/abs-1712-05889} also addresses fault-tolerance.  In general, versatile and efficient tools to improve productivity, such as easy-to-use and low-overhead monitoring and profiling of RL training are must haves.

From a computational performance viewpoint, another challenge for RL algorithms is choosing appropriate implementations for a task based on the available compute resources in order to ensure the fastest global convergence of an algorithm. Making use of resources such as multi-core CPUs, GPUs, and TPUs optimally is challenging.  Ray partially addresses this through its resource aware scheduler. It allows the user to state resource requirements, such as the number of CPUs, GPUs, or custom resources, as code annotations. This can be used to fine tune the computational performance of tasks at a high-level without the need for the user to understand or intervene in the task scheduling. 

%% file: section_uncertainty.tex
\section{Uncertainty of outcomes and insufficiency of the classical reinforcement learning theory}

In the majority of standard RL applications the agents' rewards are assumed deterministic. Contrary to this assumption, electronic trading agents typically operate in an environment where the uncertainty of outcomes is built-in. It is tempting to declare this uncertainty a ``noise'' on top of a hidden data-generating process, and it is indeed the default approximation. In the data-driven machine learning culture and in the algorithmic culture the uncertainty of outcomes is not ``noise'', it's how it all works.  We can't simply aggregate away the uncertainty of markets for it matters \emph{instrumentally.} 

As we show in other sections of the paper, the value of outcomes in electronic trading is multi-dimensional and these dimensions are often incommensurable.  Facing regulatory recommendations and restrictions and clients' instructions, we also need to have a robust way to incorporate the hierarchy of soft constraints and prohibited actions.

This inherent uncertainty of outcomes and the rich multidimensional structure of rewards challenge the standard RL theory where agents learn actions that lead to a better scalar-valued outcome \textit{on average}. In finance we, too, value aggregate outcomes, but also we value the tails of the distributions of outcomes. We need to have a methodology to combine both.

A mild extension to the standard RL methodology has been proposed: to incorporate utility functions to value multidimensional and uncertain outcomes. As in other financial applications such as portfolio construction, the agent learns good actions in the certainty equivalent sense: uncertain outcomes and their aggregates are ranked by taking the expectation of the utility function of  outcomes over their future distribution.   

Consider for example the case of a scalar uncertain reward for a finite process (to allow us to ignore the discount factor) for which the global reward is the sum of local rewards. This case reflects a typical electronic trading set up: to provide best \emph{possible} outcome on a per-share basis of the asset traded. The overall sum of rewards is still uncertain. The certainty equivalent ($\mathrm{CE}$) modification of the standard RL equation is (see also~\cite{BuehlerGononTeichmannWood2018arXiv} and~\cite{DBLP:journals/ml/MihatschN02}):
\begin{align}
	\mathrm{CE}( \pi(a_i|s_i))  &= U^{-1} \mathbb{E}\left[  U\left( r_{i+1} (\pi(a_i|s_i))+ \max_{\pi(a_{i+1}|s_{i+1})} \mathrm{CE}(\pi(a_{i+1}|s_{i+1}))  \right) \right]
\end{align}
where $U$ and $U^{-1}$ is the utility function and its inverse, $\mathbb{E}$ denotes expectation, $\mathrm{CE}$ denotes certainty equivalent: $\mathrm{CE}(\cdot) = U^{-1} \mathbb{E} \left[ U(\cdot) \right]$,  $\pi(a_i|s_i)$ is the policy $\pi$ action in the state $s_i$, and $r_{i+1} (\pi(a_i|s_i)) $ is its uncertain reward.  

The use of utility functions and certainty equivalent ranking of actions introduces a much richer agent  structure compared with that of traditional RL: in the CERL the agent acquires \emph{a character} based, however primitively,  on its risk preferences and constraints and objectives imposed by the overarching business objectives.    
If the client is risk-averse, the increased uncertainty of outcomes lowers the certainty equivalent reward of an action.  The neat consequence of this is the emergence of the discount factor $\gamma$. In classical RL it is often introduced as an exogenous parameter for infinite or nearly infinite processes.  In CERL it is naturally derived as the consequence of the broadening distribution of outcomes (an equivalent of the increased risk) as we look further into the future. 

\section{Conclusion}
Many questions remain. We hope they add new perspective to challenging problems.
\begin{itemize}[noitemsep]
	\item Is there a rigorous way to account for multidimensional rewards? 
	\item How to incorporate the concept of processes of uncertain duration into the MDP paradigm?
	\item How to tackle uncertain outcomes/rewards?
	\item How to create realistic training environments for market-operating agents? A possible solution is to develop full scale artificial environment realistically reproducing markets as \emph{emergent phenomena} arising from rule-based activities of multiple heterogeneous agents. Simulated multi-agent markets will have both practical and academic value. 
	\item How to rigorously combine conflicting/complementary local and global rewards?
	\item Other than using domain knowledge to separate processes of different time scales, and using hierarchical training, is there a rigorous way to design agents operating on multiple time scales?
	\item Scalability: in electronic trading it seems computationally efficient to train many agents operating in similar, but ultimately distinct environments, rather than one agent which is supposed to handle all environments. Is there a way for the agents trained for different environments to benefit from each other's skills? Other than testing their functionality, is there a way to tell that two trained agents are intrinsically similar?
	\item Bellman's equation in either classical RL or CERL is not fundamental and ultimately seems applicable only to processes where the global reward is a sequential aggregate of local rewards. Can a more general approach to sequential decision-making be developed which will incorporate the above characteristics?
	\item Is there a balanced and systematic approach which, on one hand, allows RL-trained agents to tackle increasingly complex problems and on the other hand, still preserves our ability to understand their behaviours and explain their actions.
\end{itemize}

%% file: main.bbl
\begin{thebibliography}{27}
\providecommand{\natexlab}[1]{#1}
\providecommand{\url}[1]{\texttt{#1}}
\expandafter\ifx\csname urlstyle\endcsname\relax
  \providecommand{\doi}[1]{doi: #1}\else
  \providecommand{\doi}{doi: \begingroup \urlstyle{rm}\Url}\fi

\bibitem[Agha and Hewitt(1987)]{DBLP:books/mit/shriverW87/AghaH87}
Gul Agha and Carl Hewitt.
\newblock Actors: {A} conceptual foundation for concurrent object-oriented
  programming.
\newblock In \emph{Research Directions in Object-Oriented Programming}, pages
  49--74. MIT, 1987.

\bibitem[Akbarzadeh et~al.(2018)Akbarzadeh, Tekin, and van~der
  Schaar]{DBLP:journals/tsp/AkbarzadehTS18}
Nima Akbarzadeh, Cem Tekin, and Mihaela van~der Schaar.
\newblock Online learning in limit order book trade execution.
\newblock \emph{{IEEE} Trans. Signal Processing}, 66\penalty0 (17):\penalty0
  4626--4641, 2018.

\bibitem[Armstrong(2010)]{DBLP:journals/cacm/Armstrong10}
Joe Armstrong.
\newblock Erlang.
\newblock \emph{Commun. {ACM}}, 53\penalty0 (9):\penalty0 68--75, 2010.
\newblock URL \url{http://doi.acm.org/10.1145/1810891.1810910}.

\bibitem[Bacon et~al.(2017)Bacon, Harb, and Precup]{DBLP:conf/aaai/BaconHP17}
Pierre{-}Luc Bacon, Jean Harb, and Doina Precup.
\newblock The option-critic architecture.
\newblock In \emph{{AAAI}}, pages 1726--1734. {AAAI} Press, 2017.
\newblock URL
  \url{https://aaai.org/ocs/index.php/AAAI/AAAI17/paper/view/14858}.

\bibitem[Bellemare et~al.(2018)Bellemare, Castro, Gelada, Kumar, and
  Moitra]{dopamine}
Marc~G. Bellemare, Pablo~Samuel Castro, Carles Gelada, Saurabh Kumar, and
  Subhodeep Moitra.
\newblock Dopamine, 2018.
\newblock URL \url{https://github.com/google/dopamine}.

\bibitem[Bergstra et~al.(2011)Bergstra, Bardenet, Bengio, and
  K{\'{e}}gl]{DBLP:conf/nips/BergstraBBK11}
James Bergstra, R{\'{e}}mi Bardenet, Yoshua Bengio, and Bal{\'{a}}zs
  K{\'{e}}gl.
\newblock Algorithms for hyper-parameter optimization.
\newblock In \emph{{NIPS}}, pages 2546--2554, 2011.
\newblock URL
  \url{https://papers.nips.cc/paper/4443-algorithms-for-hyper-parameter-optimization}.

\bibitem[{B{\"u}hler} et~al.(2018){B{\"u}hler}, Gonon, Teichmann, and
  Wood]{BuehlerGononTeichmannWood2018arXiv}
Hans {B{\"u}hler}, Lukas Gonon, Josef Teichmann, and Ben Wood.
\newblock {Deep Hedging}, 2018.
\newblock URL \url{https://arxiv.org/abs/1802.03042}.

\bibitem[Deepmind(2018)]{trfl}
Deepmind.
\newblock {TRFL}, 2018.
\newblock URL \url{https://github.com/deepmind/trfl}.

\bibitem[Dhariwal et~al.(2017)Dhariwal, Hesse, Klimov, Nichol, Plappert,
  Radford, Schulman, Sidor, Wu, and Zhokhov]{baselines}
Prafulla Dhariwal, Christopher Hesse, Oleg Klimov, Alex Nichol, Matthias
  Plappert, Alec Radford, John Schulman, Szymon Sidor, Yuhuai Wu, and Peter
  Zhokhov.
\newblock Open{AI} {B}aselines, 2017.
\newblock URL \url{https://github.com/opfenai/baselines}.

\bibitem[Doering et~al.(2017)Doering, Fairbank, and
  Markose]{doering2017convolutional}
Jonathan Doering, Michael Fairbank, and Sheri Markose.
\newblock Convolutional neural networks applied to high-frequency market
  microstructure forecasting.
\newblock In \emph{CEEC}, pages 31--36. IEEE, 2017.
\newblock URL \url{https://ieeexplore.ieee.org/document/8101595}.

\bibitem[Espeholt et~al.(2018)Espeholt, Soyer, Munos, Simonyan, Mnih, Ward,
  Doron, Firoiu, Harley, Dunning, Legg, and
  Kavukcuoglu]{DBLP:conf/icml/EspeholtSMSMWDF18}
Lasse Espeholt, Hubert Soyer, R{\'{e}}mi Munos, Karen Simonyan, Volodymyr Mnih,
  Tom Ward, Yotam Doron, Vlad Firoiu, Tim Harley, Iain Dunning, Shane Legg, and
  Koray Kavukcuoglu.
\newblock {IMPALA:} scalable distributed deep-{RL} with importance weighted
  actor-learner architectures.
\newblock In \emph{{ICML}}, volume~80, pages 1406--1415, 2018.
\newblock URL \url{http://proceedings.mlr.press/v80/espeholt18a.html}.

\bibitem[{European Securities and Market Authority}(2014)]{MiFIDII}
{European Securities and Market Authority}.
\newblock Markets in {F}inancial {I}nstruments {D}irective {II}, 2014.
\newblock URL \url{https://publications.europa.eu/s/iPhY}.

\bibitem[Fox et~al.(2017)Fox, Krishnan, Stoica, and
  Goldberg]{DBLP:journals/corr/FoxKSG17}
Roy Fox, Sanjay Krishnan, Ion Stoica, and Ken Goldberg.
\newblock Multi-level discovery of deep options, 2017.
\newblock URL \url{https://arxiv.org/abs/1703.08294}.

\bibitem[Gauci et~al.(2018)Gauci, Conti, Liang, Virochsiri, He, Kaden,
  Narayanan, and Ye]{gauci2018horizon}
Jason Gauci, Edoardo Conti, Yitao Liang, Kittipat Virochsiri, Yuchen He,
  Zachary Kaden, Vivek Narayanan, and Xiaohui Ye.
\newblock Horizon: Facebook's open source applied reinforcement learning
  platform, 2018.
\newblock URL \url{https://arxiv.org/abs/1811.00260}.

\bibitem[Kulkarni et~al.(2016)Kulkarni, Narasimhan, Saeedi, and
  Tenenbaum]{DBLP:conf/nips/KulkarniNST16}
Tejas~D. Kulkarni, Karthik Narasimhan, Ardavan Saeedi, and Josh Tenenbaum.
\newblock Hierarchical deep reinforcement learning: Integrating temporal
  abstraction and intrinsic motivation.
\newblock In \emph{{NIPS}}, pages 3675--3683, 2016.
\newblock URL \url{http://arxiv.org/abs/1604.06057}.

\bibitem[Mihatsch and Neuneier(2002)]{DBLP:journals/ml/MihatschN02}
Oliver Mihatsch and Ralph Neuneier.
\newblock Risk-sensitive reinforcement learning.
\newblock \emph{Machine Learning}, 49\penalty0 (2-3):\penalty0 267--290, 2002.
\newblock URL \url{https://doi.org/10.1023/A:1017940631555}.

\bibitem[Mnih et~al.(2013)Mnih, Kavukcuoglu, Silver, Graves, Antonoglou,
  Wierstra, and Riedmiller]{DBLP:journals/corr/MnihKSGAWR13}
Volodymyr Mnih, Koray Kavukcuoglu, David Silver, Alex Graves, Ioannis
  Antonoglou, Daan Wierstra, and Martin~A. Riedmiller.
\newblock Playing atari with deep reinforcement learning.
\newblock \emph{CoRR}, 2013.
\newblock URL \url{http://arxiv.org/abs/1312.5602}.

\bibitem[Mnih et~al.(2016)Mnih, Badia, Mirza, Graves, Lillicrap, Harley,
  Silver, and Kavukcuoglu]{DBLP:conf/icml/MnihBMGLHSK16}
Volodymyr Mnih, Adri{\`{a}}~Puigdom{\`{e}}nech Badia, Mehdi Mirza, Alex Graves,
  Timothy~P. Lillicrap, Tim Harley, David Silver, and Koray Kavukcuoglu.
\newblock Asynchronous methods for deep reinforcement learning.
\newblock In \emph{{ICML}}, volume~48, pages 1928--1937, 2016.
\newblock URL \url{http://jmlr.org/proceedings/papers/v48/mniha16.html}.

\bibitem[Moritz et~al.(2017)Moritz, Nishihara, Wang, Tumanov, Liaw, Liang,
  Paul, Jordan, and Stoica]{DBLP:journals/corr/abs-1712-05889}
Philipp Moritz, Robert Nishihara, Stephanie Wang, Alexey Tumanov, Richard Liaw,
  Eric Liang, William Paul, Michael~I. Jordan, and Ion Stoica.
\newblock Ray: {A} distributed framework for emerging {AI} applications, 2017.
\newblock URL \url{http://arxiv.org/abs/1712.05889}.

\bibitem[Nair et~al.(2015)Nair, Srinivasan, Blackwell, Alcicek, Fearon, Maria,
  Panneershelvam, Suleyman, Beattie, Petersen, Legg, Mnih, Kavukcuoglu, and
  Silver]{DBLP:journals/corr/NairSBAFMPSBPLM15}
Arun Nair, Praveen Srinivasan, Sam Blackwell, Cagdas Alcicek, Rory Fearon,
  Alessandro~De Maria, Vedavyas Panneershelvam, Mustafa Suleyman, Charles
  Beattie, Stig Petersen, Shane Legg, Volodymyr Mnih, Koray Kavukcuoglu, and
  David Silver.
\newblock Massively parallel methods for deep reinforcement learning.
\newblock \emph{CoRR}, 2015.
\newblock URL \url{http://arxiv.org/abs/1507.04296}.

\bibitem[Nevmyvaka et~al.(2006)Nevmyvaka, Feng, and
  Kearns]{DBLP:conf/icml/NevmyvakaFK06}
Yuriy Nevmyvaka, Yi~Feng, and Michael~J. Kearns.
\newblock Reinforcement learning for optimized trade execution.
\newblock In \emph{{ICML}}, pages 673--680, 2006.
\newblock URL \url{https://doi.org/10.1145/1143844.1143929}.

\bibitem[Norvig(2011)]{PN:2011}
Peter Norvig.
\newblock On {Chomsky} and the two cultures of statistical learning, 2011.
\newblock URL \url{http://norvig.com/chomsky.html}.

\bibitem[Osborne et~al.(2009)Osborne, Garnett, and
  Roberts]{osborne2009gaussian}
Michael~A Osborne, Roman Garnett, and Stephen~J Roberts.
\newblock Gaussian processes for global optimization.
\newblock In \emph{3rd international conference on learning and intelligent
  optimization (LION3)}, pages 1--15, 2009.

\bibitem[Sutton et~al.(1999)Sutton, Precup, and Singh]{sutton1999between}
Richard~S Sutton, Doina Precup, and Satinder Singh.
\newblock Between {MDP}s and semi-{MDP}s: A framework for temporal abstraction
  in reinforcement learning.
\newblock \emph{Artificial intelligence}, 112\penalty0 (1-2):\penalty0
  181--211, 1999.

\bibitem[Tian et~al.(2017)Tian, Gong, Shang, Wu, and
  Zitnick]{DBLP:conf/nips/TianGSWZ17}
Yuandong Tian, Qucheng Gong, Wenling Shang, Yuxin Wu, and C.~Lawrence Zitnick.
\newblock {ELF:} an extensive, lightweight and flexible research platform for
  real-time strategy games.
\newblock In \emph{{NIPS}}, pages 2656--2666, 2017.
\newblock URL \url{http://arxiv.org/abs/1707.01067}.

\bibitem[Vezhnevets et~al.(2017)Vezhnevets, Osindero, Schaul, Heess, Jaderberg,
  Silver, and Kavukcuoglu]{DBLP:conf/icml/VezhnevetsOSHJS17}
Alexander~Sasha Vezhnevets, Simon Osindero, Tom Schaul, Nicolas Heess, Max
  Jaderberg, David Silver, and Koray Kavukcuoglu.
\newblock Fe{U}dal networks for hierarchical reinforcement learning.
\newblock In \emph{{ICML}}, volume~70, pages 3540--3549. {PMLR}, 2017.
\newblock URL \url{http://proceedings.mlr.press/v70/vezhnevets17a.html}.

\bibitem[Zhang et~al.(2018)Zhang, Zohren, and Roberts]{zhang2018deeplob}
Zihao Zhang, Stefan Zohren, and Stephen Roberts.
\newblock Deep{LOB}: {D}eep convolutional neural networks for limit order
  books, 2018.
\newblock URL \url{https://arxiv.org/abs/1808.03668}.

\end{thebibliography}
